\documentstyle[epsf]{article}

\begin{document}
\vbox{
\begin{flushright}
 \vskip -4 cm
 \large OUTP 9407P \\
 \large 26th April 1994
\end{flushright}
\title{Multiple Ising Spins Coupled to 2d Quantum Gravity}
\author{M. G. HARRIS and J. F. WHEATER \\ \\
  \small Department of Physics\\
  \small Theoretical Physics, University of Oxford,\\
  \small 1 Keble Road, Oxford OX1 3NP, UK \\ 
  \small E-mail addresses: harris@thphys.ox.ac.uk and jfw@thphys.ox.ac.uk}
\date{}
\maketitle
\begin{abstract}
We study a model in which $p$ independent Ising spins are coupled to 2d
quantum gravity (in the form of dynamical planar $\phi^3$ graphs).
Consideration is given to the $p \! \to \!\infty$ limit in which the
partition function becomes dominated by certain graphs; we identify
most of these graphs.
A truncated model is solved exactly providing
information about the behaviour of the full model in
the limit $\beta \to 0$.
Finally, we derive a bound for the critical value of
the coupling constant, $\beta_c$ and examine the magnetization
transition in the limit $p \to 0$.
\end{abstract}}


\newcount\mycount
\mycount=1
\def\eqq{\eqno  ( \number\mycount ) \global\advance\mycount by 1}
\everydisplay{\advance\displaywidth by 2 cm}

\newcount\rra
\newcount\rrb
\newcount\eqa
\newcount\eqb
\newcount\hight
\newcount\eqasy
\newcount\eqlargn
\newcount\graphI
\newcount\graphT
\newcount\eqsum
\newcount\eqzero
\newcount\eqtriv

\advance\leftskip by - 2 cm
\advance\rightskip by -2 cm

\def\limn{\lim_{N \rightarrow \infty}}
\def\half{\frac{1}{2}}
\def\third{\frac{1}{3}}
\def\rbra{\left< r \right>}
\def\root3{\sqrt{3}}
\def\r3{{\left(\frac{r}{3}\right)}}
\def\dgdy{\frac{\partial {\cal Z}}{\partial y}}
\def\dgdx{\frac{\partial {\cal Z}}{\partial x}}
\def\sbra{\langle s \rangle}
\def\ph3{$\phi^3$}
\def\gst{\gamma_{str}}
\def\sg{{s_{\scriptscriptstyle G}}}
\def\ssg{{{\scriptstyle s}_{\scriptscriptstyle G}}}
\def\similar{\quad {\mathop\sim\limits_{\scriptscriptstyle
 N \to \infty} } \quad}
\def\og{\overline{\gamma}}

\def\gany{{\cal G}(N)}
\def\gone{{{\cal G}^{(1)}(N)}}
\def\ggone{{\cal G}^{(1)}}
\def\gtwo{{{\cal G}^{(2)}(N)}}
\def\g3{{\cal G}^{(3)}}

\def\mone{{I}}
\def\mtwo{{II}}
\def\m3{{III}}

\def\graphI{{12}}
\def\graphT{{13}}

\setlength{\unitlength}{1mm}

\section{Introduction} 
Models in which conformal matter of central charge $c$ is coupled to 2d quantum
gravity have attracted considerable interest. Much progress has been
made for the case $c \le 1$, using both continuum and discrete
approaches. The former has yielded
the KPZ formulae~\cite{KPZ,Dav1,DK}, which give as functions of the
central charge the modification of critical exponents due to gravity.
However, these formulae only apply for $c \le 1$; for larger
values of $c$ they predict that the string susceptibility
$\gst$ is complex, which does not make much sense.
The discrete methods involve using dynamical
triangulations coupled to matter fields. Matrix model techniques have
proven to be extremely valuable for studying $c\le 1$
models~\cite{BreKaz,GroMig,DouShe}
and perturbative expansions allow one to investigate numerically the
behaviour of $\gst$ for $c>1$~\cite{BreHik,HikBre,Hik}.
The model in which a single
Ising spin is attached to the face of each triangle has a central
charge of one half and has been solved
exactly~\cite{Meh,Kaz1,Kaz2,BouKaz1,BurJur}, yielding results for the
critical exponents that agree with those of the KPZ formulae.
Recently some remarkable
analytical progress has been made for the $c \! \to \! \infty$ limit by
considering the low temperature expansion~\cite{Wex1,Wex2,ADJ}.
There is evidence
that in this limit tree-like or branched polymer
graphs dominate the behaviour of the model
and this is supported by our results. Despite
much effort no one has managed to solve analytically any other models for $c
\ge 1$; however such models are well-defined and there is no obvious
pathology at $c=1$.
Many Monte Carlo simulations have been
performed~\cite{BaiJoh1,BaiJoh2,ADJT,BFHM,AmbTho,KowKrz} in an attempt
to investigate the behaviour near the $c=1$ barrier and to discover
whether the breakdown of KPZ theory is due to some change in the
geometry, such as a transition to a branched polymer phase.
So far these simulations have failed to produce any convincing
evidence of such a phase transition at $c=1$. 

In this paper we study the properties of a model for which $p$
independent Ising spins are attached to the face of each triangle,
giving a central charge of $c=p/2$. In section 2 we define precisely three
slightly different models and in section 3 we
examine the limit of large $p$, identifying most of the dominant
graphs. 
A version of the model in which the free energy is truncated is used
in section 4 to study the transition 
to behaviour typical of large $p$,
in the limit of small $\beta$. In section~\ref{sec:mag} we
study the properties of the magnetization transition,
deriving a bound
on the critical value of the coupling constant ($\beta_c \ge 0.549$)
and looking at the nature of the transition in the limit $p \to 0$.
In section~\ref{sec:conc} we conclude by discussing possible forms of
the phase diagram and by relating our work to the results from various
computer simulations and to analytical work carried out by other authors.

\section{Definition of the model} 
The model is one of $p$ independent Ising
 spins on each vertex of a random
\ph3 graph (which is the dual graph of a triangulated random
surface). For a fixed $N$-vertex \ph3 graph, $G$,
 with a single spin on each vertex,
the Ising partition function is
$$ Z_G = \frac{1}{Z_0} 
\sum_{\{S\}} \exp \left( \beta \sum_{<i j>} S_i S_j \right) \ , \eqq$$
with
$$ Z_0 = 2^N \left( \cosh \beta \right)^\frac{3N}{2}, \eqq $$
where $S_i$ is the spin on the $i$-th vertex of the graph, the sum over
$<i j>$ is a sum over nearest neighbours (by which we mean vertices
that are connected by a line in the graph) and
$\beta$ is the coupling constant. The factor of ${Z_0}^{-1}$ is
introduced in order to simplify the formulae later on.
We then sum over a set of \ph3
graphs, with $N$ vertices and a fixed genus, $g$, to get the partition
function,
\eqsum=\mycount
$$ Z_N(p)= \sum_G \frac{1}{\sg} (Z_G)^p. \eqq$$
Each graph is weighted by the symmetry factor ${\sg}^{-1}$,
which is the reciprocal of the order of the symmetry
group for that graph.
The symmetry factors are inserted because they occur in the
matrix model solution of the $p=0$ and $p=1$ cases. However, we could equally
well take $\sg=1$ in the above definition, giving us a
slightly different model, which is nonetheless expected to have
identical properties in the thermodynamic limit $N \to \infty$.
Consequently we will often ignore the symmetry
factors especially in the large $N$ limit when we would expect the
graphs not to be very symmetric and hence to have $\sg \approx 1$ anyway.

In this paper we work with planar diagrams (ie $g=0$) and
consider three different versions of this model, which differ with
respect to the sets of graphs that are used. In model~\mone, $G$ runs over all
the planar connected \ph3 graphs with $N$ vertices.
Model~\mtwo~is the same as model~\mone, except that tadpoles are
excluded so that the graphs are one-particle irreducible.
For model~\m3, tadpoles and self-energy terms are excluded
giving two-particle irreducible graphs.

In the thermodynamic limit the partition function
has the asymptotic form,
\eqasy=\mycount
$$ Z_N(p) = e^{\mu(p,\beta) N} N^{\gamma_{str} - 3}
 \left( a_0 + \frac{a_1}{N} +
\cdots \right) . \eqq$$
The free energy $\mu(p,\beta)$ is defined 
as
$$ \mu(p,\beta) = \lim_{N \rightarrow \infty} \frac{1}{N} \log(Z_N(p)) \eqq$$
and discontinuities in its derivatives indicate the presence of a
phase transition.
Similarly, $\mu_G$ for a given graph, $G$, is defined by,
$$ \mu_G(\beta) = \lim_{N \to \infty} \frac{1}{N} \log(Z_G). \eqq$$
The string exponent $\gst$ is believed to be universal and to depend
only upon certain general characteristics of the model.

\subsection{Matrix model results ($p=0$ and $p=1$)} 
The case $p=0$, where there are no Ising spins,
just corresponds to enumerating \ph3 graphs.
The number of graphs $\gone$ in model~\mone~can be calculated, for
example by matrix model methods~\cite{BIPZ} (see
also~\cite{Tutte,Koplik}) with the result,
\eqb=\mycount
$$ \gone = \frac{8^\frac{N}{2} \Gamma(\frac{3N}{4})}{2
(\frac{N}{2}+2)! \ \Gamma(\frac{N}{4} +1)} 
\similar
e^{\half \log(12 \sqrt{3}) N} N^{-\frac{7}{2}} . \eqq $$
For models~\mtwo~and~\m3, the number of $N$-vertex graphs are denoted
by $\gtwo$ and $\g3 (N)$ respectively and can also be calculated giving,
$$ \gtwo = \frac{2^\frac{N}{2} \left( \frac{3N}{2} -1 \right)!}{
\left(\frac{N}{2}\right)! \left(N+2\right)!} \similar
e^{\half \log \left( \frac{27}{2} \right) N} N^{- \frac{7}{2}}, \eqq $$
\eqa=\mycount
$$ \g3 (N) = \frac{(2N-3)!}{\left(\frac{N}{2}\right)!
\left(\frac{3N}{2}\right)!}  \similar  e^{\half \log(\frac{256}{27}) N}
N^{-\frac{7}{2}}. \eqq $$
It should be noted that all these models yield the asymptotic
form given in (\number\eqasy) with $\gst= -\half$.

The $p=1$ case has been solved analytically~\cite{Meh,Kaz1}
and has a third order
phase transition from a disordered to a magnetized state. For 
model~\mone, the critical value of the coupling constant is given 
by~\cite{BouKaz1},
$$ \beta_c = - \half \log \left( \frac{1}{27} \left( 2 \sqrt{7} -1 \right)
\right) \approx 0.9196, \eqq $$
and for model~\m3~\cite{BurJur},
$$ \beta_c = \half \log \left( \frac{108}{23} \right) \approx 0.7733 .\eqq $$
In both cases, the critical exponents are 
$\alpha = -1$ , $\beta = \half$, $\gamma = 2$,
$\delta =5$, $\nu = \frac{3}{d_H}$ and $\eta= 2 - \frac{2d_H}{3}$,
where $d_H$ is some unknown dimension depending on the geometry of the graphs.
Also, $\gamma_{str}=-\half$ everywhere, except at the critical point, 
where $\gamma^{*}_{str}=-\third$.

These results should be compared with those for a fixed regular
lattice, such as the hexagonal lattice, for which,
$$ \beta_c = \tanh^{-1}\left( \frac{1}{\root3} \right) \approx 0.658
\eqq $$
and there is a second order magnetization transition with critical
exponents: $\alpha=0$, $\beta=\frac{1}{8}$, $\gamma=\frac{7}{4}$,
$\delta=15$, $\nu=1$, $\eta=\frac{1}{4}$. Introducing the sum over
triangulations changes the universality class, but the precise nature
of the sum is not important.

\section{Dominant graphs in the limit $p \to \infty$} 
\subsection{Concavity of $\mu(p,\beta)$}
In this section we will ignore the symmetry factors, however it is
easy to show that all the results follow if we include them.
Using the Cauchy-Schwartz inequality on (\number\eqsum) it follows that,
$$ \left( Z_N \left( \frac{p+q}{2} \right) \right)^2 \leq Z_N(p)
Z_N(q) \eqq$$
and hence that $\mu(p,\beta)$ is concave with respect to p,
$$\mu\left( \frac{p+q}{2}, \beta \right) \leq \frac{1}{2}
  \left[ \mu\left(p,\beta\right) + \mu\left(q,\beta\right) \right]. \eqq $$
(A slight modification of the standard argument shows that $\mu$ is
also concave with respect to $\beta$).
It is trivial that for $p \geq 1$,
$$ Z_N(p) = \sum_G (Z_G)^p \leq \left( \sum_G Z_G \right)^p , \eqq $$
so that,
$$ \mu(p,\beta) \leq p \mu(1,\beta), \qquad  p \ge 1.\eqq $$
We also have that,
$$
  {\partial\mu(p,\beta) \over \partial p} = \limn \frac{1}{N} \frac{1}{Z_N}
  {\partial \over \partial p} \left( \sum_G (Z_G)^p \right) \eqq $$
$$
   \qquad = \limn \frac{1}{Z_N} \sum_G (Z_G)^p \mu_G > 0 
$$
and hence that $\mu(p,\beta)$ is a concave monotonic increasing function
with respect to $p$.
However, $\mu$ is bounded by a linear function
and hence it must be asymptotically linear in $p$ as $p \to \infty$.
Since the number of graphs with given $Z_G$ does not depend upon $p$
it must be the case that, as $p \to \infty$, $Z_N(p)$ is dominated by those
graphs $G_0$ with largest $Z_G$ so that,
$$ \mu(p \rightarrow \infty) \sim p \mu_{G_0}  \eqq $$
and it is therefore of some interest to identify these maximal graphs.
This identification alone is not sufficient to solve the models at
large $p$ because the number of such graphs and fluctuations around
them must be included to find the sub-asymptotic behaviour and, in
particular, the value of $\gst$.

In the following section we identify most of the maximal graphs for
models~\mone, \mtwo~and~\m3. For a given value of $\beta$ we might
suppose that there is a value $p_c(\beta)$ of $p$ above which the
maximal graphs dominate. (Of course it may be that the transition to
dominance by maximal graphs is seamless and that there is no critical
value or that $p_c(\beta)=\infty$ for all $\beta$.)
In section~\ref{sec:trunc} we begin to address the question of
the behaviour of $p_c(\beta)$.

\subsection{Model~\mone} 
\begin{figure}[b]
\caption[l]{Tree-like graph}
\label{fig:tree}
\begin{picture}(100,35)(-35,0)
\epsfbox{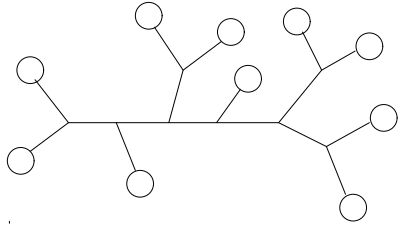}
\end{picture}
\end{figure}

In this section we prove that the maximal graphs for model~\mone~are
tree graphs with tadpoles at the ends of the branches (see
fig~\ref{fig:tree} for an example).

\begin{figure}[htb]
\caption{Partition functions}
\label{fig:treeblob}
\begin{picture}(100,45)(-10,0)
\epsfbox{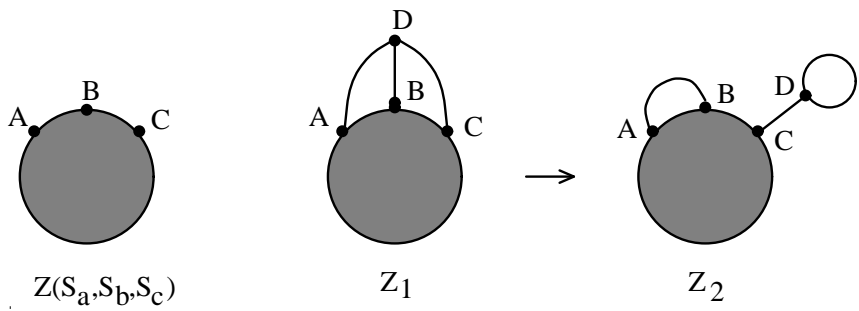}
\end{picture}
\end{figure}

Starting with an arbitrary $N$ vertex \ph3 graph, select a point D and
reconnect the links to it as shown in fig~\ref{fig:treeblob}. The
resulting graph still has $N$ vertices, and provided that the link 
DC lay on a closed loop in the original graph, the new graph will still be 
connected. Repeat this procedure taking care not to disconnect the graph.
Eventually there are no further links that can be cut and the 
graph is tree-like (ie there are no closed loops except for the 
tadpoles at the ends of the branches).
Before the link DC is cut the partition function is

\setcounter{equation}{\mycount}
\addtocounter{equation}{-1}
\global\advance\mycount by 2
\vbox{
\begin{eqnarray}
 Z_1 &=& \frac{1}{2 C^3} 
\sum_{S_a S_b S_c S_d} Z(S_a,S_b,S_c) \, \exp\beta(S_a S_d+S_b
S_d+S_c S_d)\\
     &=& \sum_{S_a S_b S_c} Z(S_a,S_b,S_c) \, (1 + t^2 (S_a S_b+S_b
S_c+S_c S_a)),
\end{eqnarray}}

\noindent where $Z(S_a,S_b,S_c)$~$(\ge 0)$ represents the partition
function for the remainder of the graph with boundary spins 
$S_a$, $S_b$, $S_c$ and $C= \cosh \beta$, $t=\tanh \beta$. After
reconnecting the links, the new partition function is
$$ Z_2  =\sum_{S_a S_b S_c} Z(S_a,S_b,S_c) \, (1 + t)(1+t S_a S_b)
\eqq$$
so that,
$$ Z_2 - Z_1 = \sum_{S_a S_b S_c} Z(S_a,S_b,S_c) \, t (1+S_a S_b)
(1-t S_b S_c) \ge 0 .\eqq $$
Thus at each step the partition function increases
(equality occurs for $\beta =0$ and $\beta = \infty$). The partition
function takes the same value,
$$ Z_{tree}  =  
 (1+\tanh \beta)^{\frac{N}{2}+1} \eqq $$
for all tree graphs with $N$ vertices, so we have proved that the set
of tree graphs is maximal for all finite non-zero $\beta$. This set of
graphs does not magnetize at finite $\beta$. We will need later
the number of tree graphs with $N$ vertices which is given by~\cite{BKKM}
$${\cal G}_{tree}(N) = \frac{(N-2)!}{(\half N +1)! (\half N -1)!}
\similar
e^{N \log 2} N^{-\frac{5}{2}} . \eqq$$

In considering the contributions of different graphs to $Z_N(p)$ at
finite $p$ (which we will do in sections~\ref{sec:mag}
 and~\ref{sec:conc}) it is useful
to examine the ratio,
$$\frac{Z_2}{Z_1} = \left(1 - \frac{t}{(1+t)^2} \left< 1 + S_a S_b - t
(S_b S_c +S_a S_c) \right>_{G_2} \right)^{-1}, \eqq$$
where
$$ \left< Q \right>_{G_2} \equiv \frac{1}{Z_2} \sum_{S_a S_b S_c}
Z(S_a,S_b,S_c) \, (1+t) (1+t S_a S_b) Q .\eqq$$
For small $t$, $\left< S_a S_b \right>=1$ or is of order $t$ (depending
on whether or not A and B are distinct points), while $\left<S_b S_c
\right>$, $\left< S_a S_c \right> \sim O(t^m)$ with $m \ge 1$; thus
$Z_2/Z_1$ increases with $t$ at small $t$ so that tree-like graphs are
becoming more important. Assuming that $p_c(\beta)$ is finite,
it is decreasing with $\beta$ in this region. On the other hand,
for large enough $t$, $\left< S_a S_b \right>$, $\left< S_b
S_c \right>$,
$\left< S_a S_c \right> \approx 1$ and $Z_2/Z_1$ decreases towards one
as $t \to 1$, so that tree-like graphs become less important again.
The position of any minimum of the curve $p_c(\beta)$ must lie between
these two regimes.

\subsection{Model~\mtwo} 
\label{sect:ring}
\begin{figure}[htbp]
\caption{Ring graph}
\label{fig:ring}
\begin{picture}(100,40)(-40,0)
\epsfbox{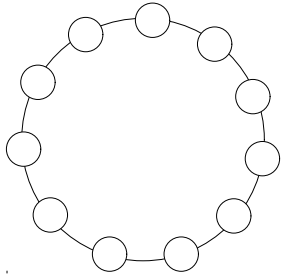}
\end{picture}
\end{figure}

For model~\mtwo, the maximal graphs are ring graphs
(fig~\ref{fig:ring}) and again this is true for any value of $\beta$.
The proof, which is very
similar to that for model~\mone, consists of two parts. Firstly, we
show that any graph in this model can be converted into a ring graph
by a series of steps, where none of the intermediate graphs
contain tadpoles. Secondly, we show that the partition function
increases at each step and thus that ring graphs are maximal for all
$\beta$.

For \ph3 graphs of genus zero the number of faces, $F$, is related to
the number of edges, $E$, through $F=2+\third E$.
Defining $f_i$ to be the number of faces with
$i$ sides, then $F=\sum f_i$ and $E=\half \sum i f_i$ so that,
$$ \sum_i f_i (1-\frac{i}{6})=2. \eqq $$
Since $f_i \ge 0$, in order for the equation to be satisfied we must
have $f_i \ne 0$ for some $i < 6$. Thus, any \ph3 graph in this model
must contain a 2-loop (ie a loop of length two), a triangle,
a square or a pentagon.

\begin{figure}[bhp]
\caption{(a) Dressed (b) Bare (c) Renormalized propagators}
\label{fig:dress}
\begin{picture}(170,45)(-5,0)
\epsfbox{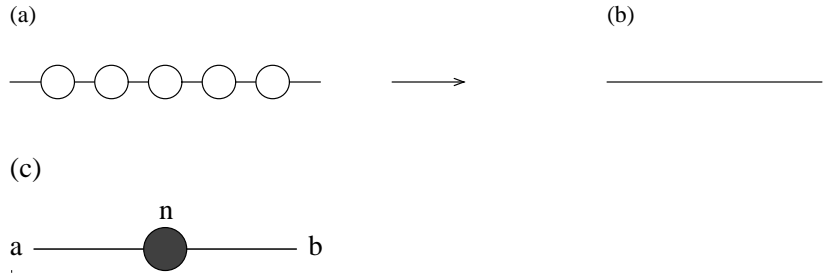}
\end{picture}
\end{figure}

Starting with a graph $G$, containing no tadpoles, replace any
subgraphs such as that in figure~\ref{fig:dress}a (which we will call
dressed propagators) with bare propagators (fig~\ref{fig:dress}b),
yielding a new \ph3 graph, $G'$, to which the above
theorem applies. Putting the dressed propagators back we recover $G$ and have
shown that it contains at least one of the following: a dressed
2-loop, or a dressed or bare triangle, square or pentagon. The only
exception is the ring graph, which upon making the replacement shown
in fig~\ref{fig:dress} just gives a circle for graph $G'$. Thus, any
graph, except for the ring, contains one of the subgraphs in the above
list.

\begin{figure}[pt]
\caption{(a) 2-loop (b) Replacement subgraph or equivalently (c)}
\label{fig:loop}
\begin{picture}(170,35)(5,0)
\epsfbox{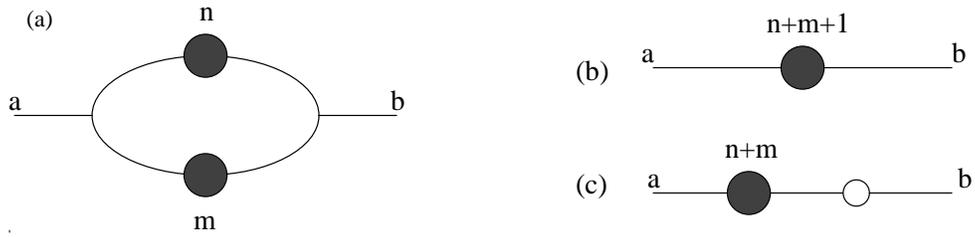}
\end{picture}
\end{figure}

Dressed propagators containing $n$ 2-loops will be drawn as in
fig~\ref{fig:dress}c. We are going to replace the subgraphs
fig~\ref{fig:loop}a to fig~\ref{fig:pent}a with subgraphs
fig~\ref{fig:loop}b to fig~\ref{fig:pent}b respectively. The number of
vertices is unchanged and the number of 2-loops is increased by these
replacements. In appendix~\ref{app:ring} we show that the partition
function increases,
for any choice of dressed propagators in the original subgraph (ie for
all choices of $j,k,l,m,n \ge 0$; for the 2-loop case,
fig~\ref{fig:loop}a, $n,m$ are not both zero).
Thus by choosing the orientation of the
replacement, we can create a new graph, which is connected,
has no tadpoles, has the same number of vertices as the original and
has a larger partition function. By repeatedly
eliminating the subgraphs in the above list
we will eventually end up with a ring
graph,
proving that the ring graphs are maximal for all $\beta$. The
partition function for the ring graph is given by,
$$ Z_{ring} = \left( 1+t^2
\right)^\frac{N}{2}  + \left( 2 t^2 \right)^\frac{N}{2} \eqq $$
and this graph does not magnetize for any finite $\beta$.

\begin{figure}[ph]
\caption{(a) Triangle (b) Replacement subgraph}
\label{fig:tri}
\begin{picture}(170,55)(5,0)
\epsfbox{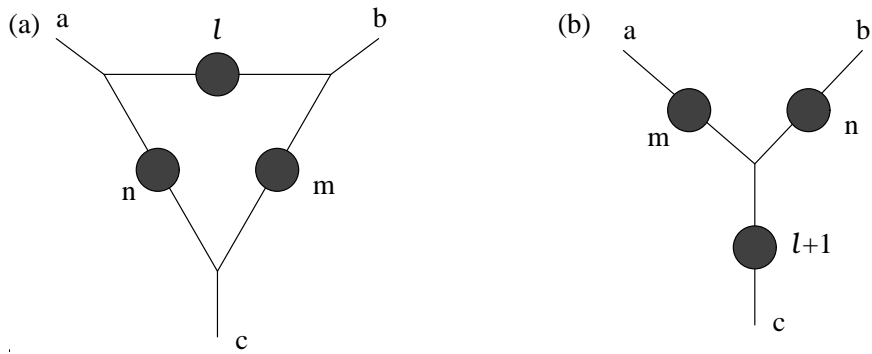}
\end{picture}
\end{figure}

\begin{figure}[pht]
\caption{(a) Square (b) Replacement subgraph}
\label{fig:squ}
\begin{picture}(170,50)(5,0)
\epsfbox{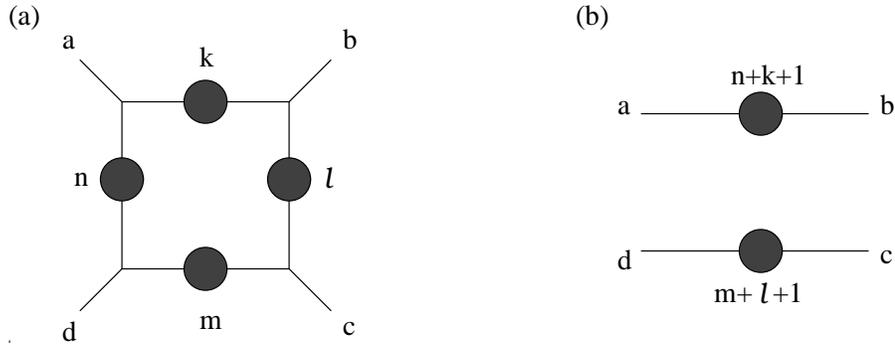}
\end{picture}
\end{figure}

\begin{figure}[pht]
\caption{(a) Pentagon (b) Replacement subgraph}
\label{fig:pent}
\begin{picture}(170,65)(5,0)
\epsfbox{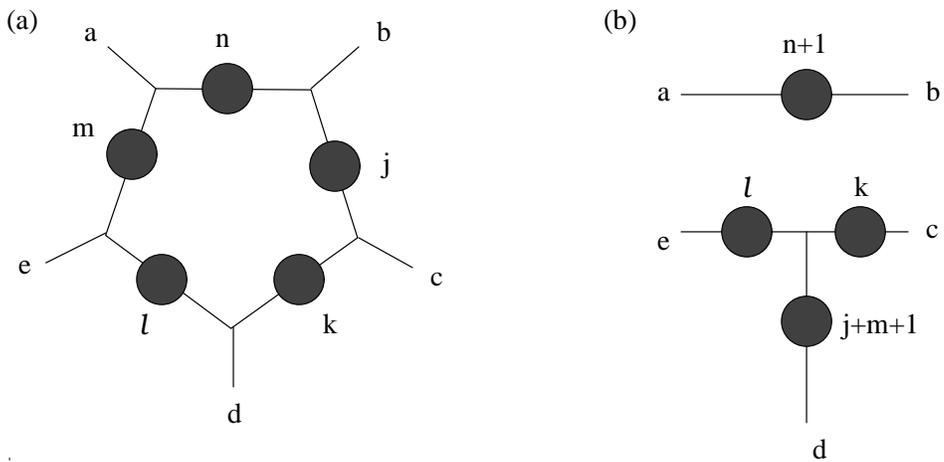}
\end{picture}
\end{figure}

\break

\subsection{Model~\m3} 
In this case we have not found a procedure which converts any given
graph into a maximal graph; essentially this is because the form of
the maximal graph now depends on $\beta$.
We have identified the maximal graphs for the limits
of small and large $\beta$, but not for intermediate values of $\beta$.

\begin{figure}[htb]
\caption{(a) Replacement (b) Tetrahedron}
\label{fig:fractet}
\begin{picture}(100,30)(0,0)
\epsfbox{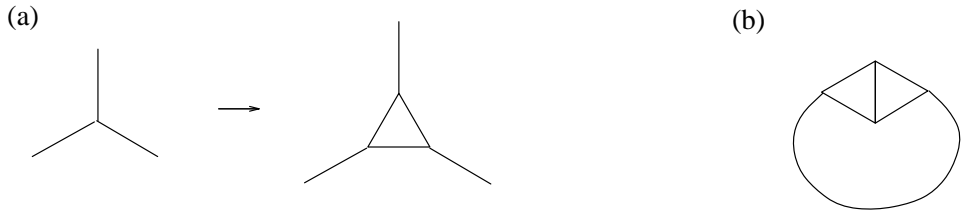}
\end{picture}
\end{figure}

A high temperature expansion $(\beta
\rightarrow 0)$ of the partition
function for a given graph gives,
\hight=\mycount
$$ Z_G = 1 + \sum_l n_l t^l  , \eqq $$
where $n_l$ is the number of closed, but possibly disconnected,
non-self-intersecting loops in the graph that contain $l$ links.
For model~\m3, $n_1=n_2=0$
because we have eliminated tadpoles and self-energies and the first
non-zero coefficient is $n_3$, the number of triangular loops.
The maximum possible value of $n_3$ is the integer part of
$\frac{N}{3}$ which we will denote by 
$\left[\frac{N}{3}\right]$. Suppose that $N$ is divisible by
three, then taking any graph with $\third N$ points and replacing
each point with a triangle (fig~\ref{fig:fractet}a), 
gives an $N$ vertex graph with $n_3=
\frac{N}{3}$. If $N$ is not divisible by three, replace all except one or
two of the points giving $n_3=\left[\frac{N}{3}\right]$. To
show that this really is the maximum possible number,
first note that the only case
which has two triangles back to back is the tetrahedron
(fig~\ref{fig:fractet}b).
Then, ignoring this case, given an $N$ vertex graph we can collapse all
of its triangular loops to points and get a
graph with $N'$ vertices and no tadpoles or self-energies.
Clearly $N' \geq n_3$ (since each collapsed triangle yields a vertex)
and $N' = N - 2 n_3$ (as collapsing a triangle removes two vertices). 
Thus $N-2n_3 \geq n_3$
and so $n_3 \leq \frac{N}{3}$. Hence the maximum value is
$n_3 = \left[\frac{N}{3}\right]$ for $N>4$.

\begin{figure}[htb]
\caption{Fractal graph}
\label{fig:fractal}
\begin{picture}(100,50)(-35,0)
\epsfbox{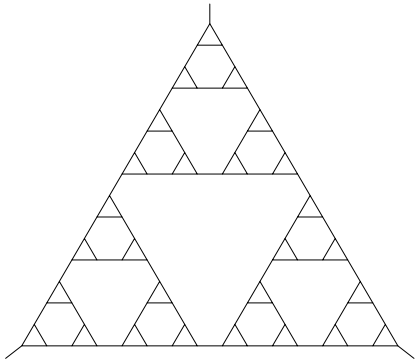}
\end{picture}
\end{figure}

Consider graphs, $G$, for which the total
number of points is $N=2 \times 3^n$, where $n$ is some sufficiently
large integer. In order to maximize $Z_G$ in the $\beta \to 0$ limit,
we maximize each coefficient in turn. Having maximized $n_3$ as
described above by replacing each point of $G'$ (with $\third N$
points) with a triangle we now choose $G'$ in order to
maximize the next coefficients. The replacement of points with
triangles doubles the
number of edges bordering each face of $G'$, which being a
model~\m3~graph has a smallest loop length of three. Any such loops
will be doubled to length six; thus $n_4=n_5=0$. Now, $n_6 = n_6^c +
\half n_3(n_3-1)$ (where $n_6^c$ is the number of connected loops of
length six), so that we need to maximize $n_6^c$ next.
To do this we must maximize $n_3$ of $G'$ (since loops of length three
in $G'$ become those of length six in $G$). So we take a graph with
$\frac{1}{9} N$ points and make the replacement of points with
triangles to get $G'$. Carrying
on in this fashion we end up with a fractal graph (see fig~\ref{fig:fractal}).
If $\half N$ is not a power of three, then the graph will not quite be
regular, but will still be essentially fractal-like.

\begin{figure}[htb]
\caption{Ladder graph}
\label{fig:ladder}
\begin{picture}(100,35)(-25,0)
\epsfbox{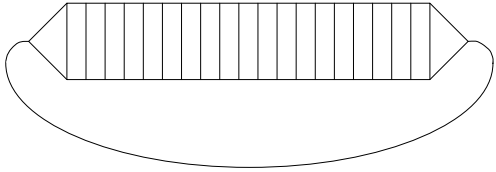}
\end{picture}
\end{figure}

At large $\beta$, graphs which magnetize can be studied in the low
temperature expansion for which,

$$ Z_G = \frac{1}{Z_0}
2 e^{\frac{3N}{2} \beta} \left( 1 + \sum_{r=3}^{\infty} m_r
x^r \right) \sim \frac{1}{Z_0}
\exp N \left(\frac{3}{2}\beta + \sum_{s=3}^{\infty}
a_s x^s \right) , \eqq$$
where $x=e^{-2 \beta}$ and $m_r$ is the number of
domain boundaries that cross $r$ links.
For graphs with no tadpoles or self-energies any
domain boundary must cross at least three links so that the sum starts
at $r=3$ and the sum for the free energy also starts at $s=3$.
However, this does not apply to graphs that do not magnetize; the
ladder graph (fig~\ref{fig:ladder}) has $m_3 \propto N$, but $m_4 =
\half \frac{N}{2} \left( \frac{N}{2} -1 \right) $ so we might expect
it to exponentiate to give a series starting at $s=2$ with $a_2=
\half$. It is straightforward to check that this is the case from the
ladder free energy, which is given by,
$$ \mu_{ladder} = \frac{1}{2} \log \left( \half \left(1 + t^2 +
\sqrt{(1- t^2)^2 + 4 t^4}\right) \right) . \eqq $$
Because these graphs are one-particle irreducible it is not possible
to get $m_2 \propto N^2$ and $a_1$ is always zero. We conclude that at
large $\beta$ the ladder graphs, whose free energy starts at $O(x^2)$,
will dominate magnetizable graphs, whose free energy starts at
$O(x^3)$. Furthermore it is easy to check that $\mu_{ladder}$ is
bigger than that of the fractal graph for large $\beta$.

What happens at intermediate $\beta$ is not clear but is probably
quite complicated, involving graphs which in some sense interpolate
between the fractal and the ladder graphs; whether the transition from
fractal to ladder is continuous or discontinuous we cannot say, but
the former seems more likely on grounds of universality with
models~\mone~and~\mtwo~- but maybe this system is not universal.
Neither the fractal nor the ladder magnetize so probably the
intermediate graphs do not magnetize either and as $p \to \infty$ the
model only magnetizes at $\beta \to \infty$.

It is interesting to note that for a Gaussian model on a random
triangulation embedded in $D$
dimensions~\cite{BKKM,KKM,ADFO,David,ADF3} the maximal graphs (for $D>0$)
are those with the minimal number of spanning trees and that the
corresponding \ph3 graphs are tree graphs, ring graphs and ladders for
models~{\mone} to~{\m3} respectively.
Thus the maximal Ising graphs at
large $\beta$ are the same as the maximal graphs for the Gaussian
model, but they apparently differ at small $\beta$ in the case of model~\m3.
Reference~\cite{BKKM} lists the maximal graphs, but gives a more symmetric
version of the ladder
graph; presumably that is because
fig~\ref{fig:ladder} is excluded from the relevant model even though it has no
tadpoles or self-energies.

\section{Truncated models} 
\label{sec:trunc}
\subsection{Model~\mone} 
To get some indication of how large $p$ must be before the maximal
graphs become dominant we now consider a truncated model in the weak
coupling regime. For small $\beta$,
$$\mu_G =
\frac{1}{N} \left[ n_1 t + \left( n_2 - \half {n_1}^2\right) t^2 +
\left(n_3 - n_1 n_2 +
\third {n_1}^3 \right) t^3 + \cdots \right] , \eqq $$
so we consider a model for which $\mu_G$ is truncated to
$$\mu_G^T =\frac{n_1}{N} t . \eqq $$
The grand canonical partition function for this model is then,
$$ {\cal Z}(\mu,pt) = \sum_{N=2 \atop even}^{\infty} e^{- \mu N}
\sum_{n_1 =0}^{\infty} \ggone(N,n_1) e^{n_1 p t}, \eqq$$
where $\ggone(N,n_1)$ is the number of graphs with $N$ vertices and
$n_1$ tadpoles (loops of length one) \break ($0 \le n_1 \le \half N +1$). We
show in appendix~\ref{app:tad} that $\ggone(N,n_1)$ satisfies the
recurrence relation,
\rrb=\mycount
$$ \ggone (N+2,n_1+1) =  \left( \frac{3N - 2 n_1}{n_1+1}
\right) \ggone (N,n_1)+ 2 \ggone (N,n_1+1)  \eqq $$
and the total number of graphs is known so
$$\sum_{n_1=0}^{\half N +1} \ggone (N,n_1) = \gone, \eqq $$
where $\gone$ is given in (\number\eqb). Using the recurrence relation
(\number\rrb) and putting $y=e^{pt}$, $x=e^{-\mu}$ we find that $\cal Z$
satisfies the differential equation,
$$ \dgdy (1 + 2x^2 (y -1)) = 3 x^3 \dgdx + x^2 y \ , \eqq $$
which has the solution,
$$ {\cal Z} = \frac{1}{12 x^2} \left( h^{\frac{3}{2}} -
1 \right) + \half (y-1) - \frac{1}{4} \log h +
\sum_{ N = 2 \atop even}^{\infty} \gone x^N h^{- \frac{3N}{4}} , \eqq $$
where $h = (1- 4x^2 (y-1))$. The asymptotic behaviour at large $N$ is
thus given by
\eqlargn=\mycount
$${\cal Z} \sim \sum_{N} \gone x^N h^{-\frac{3N}{4}} \sim
 \sum_{N}  N^{-\frac{7}{2}} \exp -N
\left( \mu  + \frac{3}{4} \log h  - \half \log(12 \root3) \right), \eqq $$
so the thermodynamic free energy $\mu_c(pt)$ obeys the cubic equation,
$$ \mu_c + \frac{3}{4} \log \left(1-4 e^{-2\mu_c} (y-1) \right)
 - \half \log (12 \root3) =0. \eqq $$
The solution is

\vbox{
$$ \mu_c(pt) = \log 2 + \half \log (y-1) - \half \log \Biggl[ 1 -
\frac{9}{(y-1)^2} - 3 \left( \frac{1}{2 (y-1)^2} \right)^\third \times $$
$$ \sum_{\sigma = \pm 1} \omega^{\sigma}
\left( 1- \frac{18}{(y-1)^2} +
\frac{54}{(y-1)^4} + \sigma \sqrt{1 - \frac{4}{(y-1)^2}}
\right)^\third \Biggr] , \eqq $$}
\noindent with $\omega = e^{\frac{2 \pi i}{3}}$ for $y<3$
and $\omega =1$ for $y>3$ (the solution and its derivatives are
actually continuous across $y=3$). The solution is plotted in
fig~{\graphI}a where it is compared to ${\mu_c}' = \half
pt + \log 2$, which is what we would expect if trees were totally dominant. We
see that $\mu_c(pt)$ approaches ${\mu_c}'$ quite rapidly as
$pt$ increases, but it is necessary to have $p \to \infty$ for the
trees to dominate completely. It is also interesting to examine the
ratio $r \equiv { \langle n_1 \rangle}/{N}$, which is shown in
fig~{\graphI}b; for $pt >6$ the ratio is very close to one half and
the fact that the gradient changes very quickly near this point leads
us to suspect that the full model might develop a discontinuity in one
of its derivatives at some finite value of $p$ (ie $p_c(\beta)$).
A formula for $r$ can be derived from that for $\cal Z$;
putting $b = 2 e^{- p t}$,
$\lambda = \frac{4 b^2}{(3b-2) (b+2)}$,
$$ r = \frac{1}{2-b} \left[ 1 + \lambda^\third \left( \omega \left(1 -
\sqrt{1- \lambda} \right)^\third + \omega^2 \left(1 + \sqrt{1- \lambda}
\right)^\third \right) \right] , \eqq $$
where $\omega$ is a cube root of unity and we need to use $\omega=1$
for $0 \leq b < \frac{2}{3}$ and
$\omega = e^{\frac{2 \pi}{3} i}$ for $\frac{2}{3} < b \leq 2$
(the solution and derivatives are continuous across $b=\frac{2}{3}$).

As is shown by (\number\eqlargn) the string susceptibility
$\gst=-\half$ for all finite $p$ in this truncated model; only for
$p=\infty$ does $\gst$ change to $\half$, but, again, in the full
model this change may occur at some finite $p_c(\beta)$.

\begin{figure}[hptb]
\caption{Truncated model (model I)}
\begin{picture}(200,110)(40,100)
\epsfbox{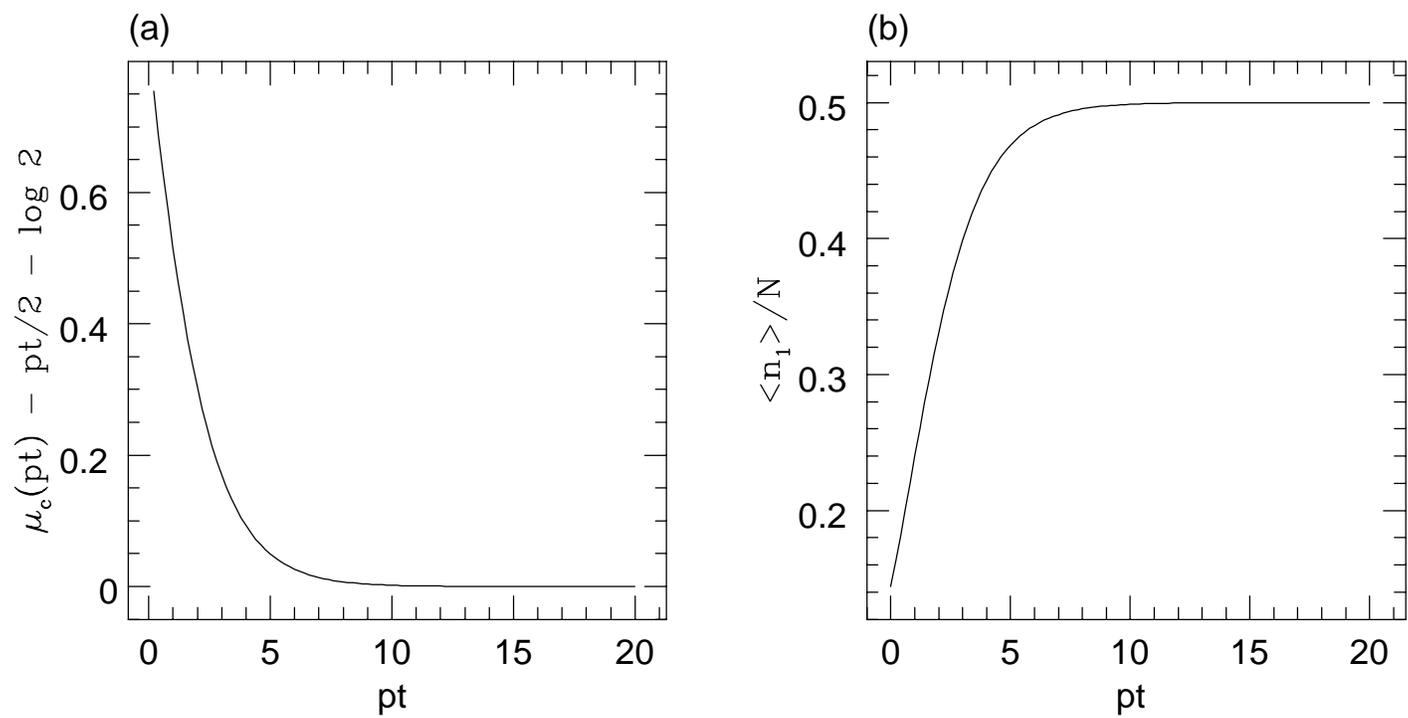}
\end{picture}
\end{figure}

\subsection{Model~\m3} 

This calculation can be repeated for model~\m3, which has,
$$ \mu_G = \frac{1}{N} (n_3 t^3 +
n_4 t^4 + n_5 t^5 + (n_6 - \half n_3^2 ) t^6 + \cdots ) . \eqq $$
We truncate $\mu_G$ to
$$ \mu_G^T =  \frac{n_3}{N} t^3 \eqq $$
where $0 \le n_3 \le \left[\frac{N}{3}\right]$.
The grand canonical partition function is defined as
$$ {\cal Z}(\mu,pt^3) = \sum_{N=6 \atop even}^{\infty} e^{-\mu N}
\sum_{n_3 =0}^{\infty} \g3(N,n_3) e^{ n_3 p t^3}, \eqq $$
where $\g3(N,n_3)$ is the number of graphs in model~{\m3} with $N$
vertices and $n_3$ loops of length three. This obeys a recurrence
relation, derived in appendix~\ref{app:tri},
\rra=\mycount
$$ \g3 (N+2,n_3 +1)= \left( \frac{N-3n_3}{n_3+1} \right) \g3 (N,n_3) +
 3 \g3 (N,n_3+1), \eqq $$
which holds for $N \geq 6$.
The recurrence relation (\number\rra) gives,
putting $y=e^{pt^3}$ and $x=e^{-\mu}$,
$$ \frac{\partial {\cal Z}}{\partial y} \left(1 + 3 x^2 \left( y-1
\right) \right) = x^3 \frac{\partial {\cal Z}}{\partial x} + \third
x^6 y . \eqq $$
The solution is
$$ {\cal Z}(\mu,pt^3) = - \third \left( h x - \half x^2 + \frac{1}{4} x^4
\right) + \sum_{N=2 \atop even}^{\infty} \g3 (N) h^N , \eqq $$
where $\g3(N)$ is given by (\number\eqa) and
$ h = x + x^3 (y-1) .$
Thus for large $N$,
$$ {\cal Z} \sim \sum_N \g3 (N) h^N \sim \sum_N N^{-\frac{7}{2}} \exp N \left( 
\half \log \left(
\frac{256}{27} \right) + \log \left( x + x^3 (y-1) \right) \right).
\eqq $$
Solving for $\mu_c(pt^3)$ as in the previous case gives
$$ \mu_c(pt^3) =  - \half \log 3 + \frac{5}{3}\log 2 + \third \log
(y-1) - \log \left[\sum_{\sigma= \pm 1}
 \left( 1 +  \sigma \sqrt{1 + \left( \frac{32}{27} \right)^2
\frac{1}{y-1} } \right)^\third \right] . \eqq $$
We can also calculate $\langle n_3 \rangle/N$,
$$ \frac{\langle n_3 \rangle}{N} = 
\frac{1}{a} \left[ 1 - \left( \half \left( \frac{3 -
a}{12 - c a} \right) \right)^\third \sum_{\sigma = \pm 1}
\left( 1+ \sigma \sqrt{1- 4 \left( \frac{3 - a}{12 - c a}
\right) } \right)^\third \right] , \eqq $$
where $ a = 3(1-e^{-pt^3}) $ and $c=295/256$.
Again, this saturates as $p$ is increased (see fig~{\graphT}b).

Looking at the graph (fig~{\graphT}a) of $\mu_c(pt^3)- \third pt^3 -
\frac{1}{6} \log\left(\frac{256}{27}\right)$ (where the last term
takes account of the exponential number of graphs with
$\frac{n_3}{N}=\third$) we see that for $p t^3 > 8$
the behaviour is essentially linear, because after this point
the integral is being dominated by a single type of graph (ie those
graphs with $n_3 \approx \third N$). However, it should be noted that in
this truncated model we again
need $p=\infty$ in order to actually reach $ n_3/N =\third$ and the set
of such graphs is quite large
(not just fractal-like graphs).
This is due to the fact that $\mu_G$ has been truncated to the
first term ($n_3$), but higher order terms are needed to show that fractals
are maximal for $\beta \to 0$. Again, $\gst =-\half$ for all
finite values of $p$.

\begin{figure}[hp]
\caption{Truncated model (model III)}
\begin{picture}(200,110)(40,0)
\epsfbox{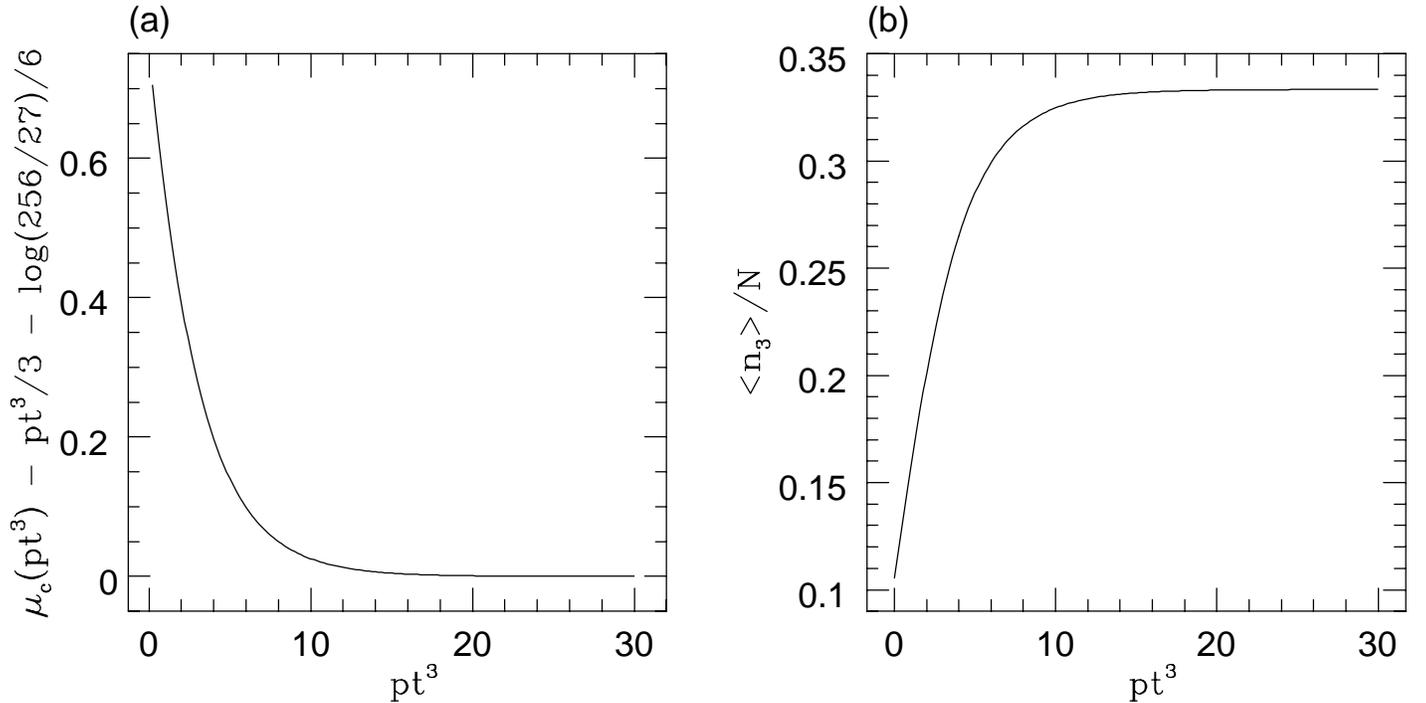}
\end{picture}
\end{figure}

\section{Magnetization transition}  
\label{sec:mag}
\subsection{Derivation of a bound on $\beta_c$}
In this section we derive a bound on the critical value
of the coupling constant, $\beta_c$, for the case of a single Ising
spin ($p=1$) on a fixed
graph $G$. This implies a bound on the critical coupling for
magnetization related phase transitions in models~\mone, \mtwo~and~\m3.
The high temperature expansion of $Z_G$ (\number\hight)
gives,
$$ Z_G =  1+ \sum_l n_l t^l  =
\exp \left({ \sum_{l=1}^{\infty} a_l t^l} \right) . \eqq $$
Now consider,
$$ Z_{G'} =  \exp \left( \sum_{l=1}^\infty n_l^c t^l \right), \eqq $$
where $n_l^c$ is the number of connected,
non-backtracking closed loops of length $l$.
Expanding the exponential yields all closed non-intersecting loops,
but in addition it gives loops in which some links are used more than
once so that $Z_G < Z_{G'}$ and therefore
$$ \mu_G <  \frac{1}{N} \sum_{l=1}^{\infty}n_l^c t^l .\eqq$$
However the number of non-backtracking closed loops of length $l$
originating at a given point is less than $2^l$ so
$$ \mu_G < \sum_{l=1}^\infty (2t)^l . \eqq$$
Thus the high temperature series for $\mu_G$ must converge if
$t<\half$; it follows that any phase transition on $G$ must occur at a
$t_c \ge \half$ (ie $\beta_c \ge 0.549$).

A slight modification of
this argument excludes the possibility of permanently magnetized graphs.
Fix one spin $S_+$ to be $+1$, so the magnetization $M_G(\beta)$ is given by,
$$ M_G(\beta) = \frac{1}{N} \sum_{+}
\frac{1}{N} \frac{1}{Z_G} \frac{1}{2^{N-1}} \sum_{\{S\}} \sum_i
S_i \prod_{\langle a,b \rangle} ( 1 + t
S_a S_b) , \eqq $$
where the $\sum\limits_{+}$ runs over all possible locations of the
fixed spin (this is necessary because unlike the regular lattice model
any two given spins may be almost disconnected from one another).
The only contribution to the numerator comes from paths connecting
$S_i$ to $S_+$, so
$$ M_G(\beta) = \frac{1}{N^2} \ \frac{ \sum\limits_{+} 2^{N-1}
 \sum\limits_i \sum\limits_l d_l(i,+)t^l}
{\sum\limits_{\{S\}} 
\prod\limits_{\langle a,b \rangle} (1+t S_a S_b)} , \eqq $$
where $d_l(i,+)$ is the number of paths of length $l$ from $S_i$ to $S_+$
(which can be disconnected, but are non-self-intersecting and
non-backtracking). However,
$$\sum_l d_l(i,+) t^l = \sum_l w_l(i,+) t^l ( 1 + b_1 t + b_2 t^2 + \cdots
) , \eqq $$
where $w_l(i,+)$ is the number of connected paths of length $l$ from $S_i$
to $S_+$ and the $b_l$ series gives the contributions from the closed
loops, which do not intercept the path. The denominator, however,
contains contributions from all closed loops and hence is larger than
$2^{N-1} \sum_l b_l t^l$ so
$$ M_G(\beta)  < \frac{1}{N^2} \sum_+\sum_i \sum_l w_l(i,+) t^l .\eqq $$
However, $\sum_i w_l(i,+)$ is just the number of connected paths of length $l$
from $S_+$ and we know that this is less than $2^l$. Thus,
$$ M_G(\beta) < \frac{1}{N} \sum_l (2t)^l = \frac{1}{N}
\frac{1}{1-2t} , \eqq $$
for $t<\half$ and hence $ M_G(\beta) \to 0$ as $N \to \infty$.
If more than one spin is fixed to be $+1$, then the
right hand side of this equation is multiplied by the number of such
spins, so that to avoid $ M_G(\beta) \to 0$ a number of spins
proportional to $N$ must be fixed. That is, we need to fix a thermodynamically
significant number of spins in order to get the system to magnetize
and thus there can be no spontaneously magnetized state for $t<\half$.

This result was derived for a single graph, but obviously still applies
when we perform a summation over graphs; no phenomena associated with
magnetization can occur at $t<\half$.

\begin{figure}[htb]
\caption{Triangulations: (a) never magnetizes; (b) permanently magnetized}
\label{fig:dual}
\begin{picture}(100,45)(22,0)
\epsfbox{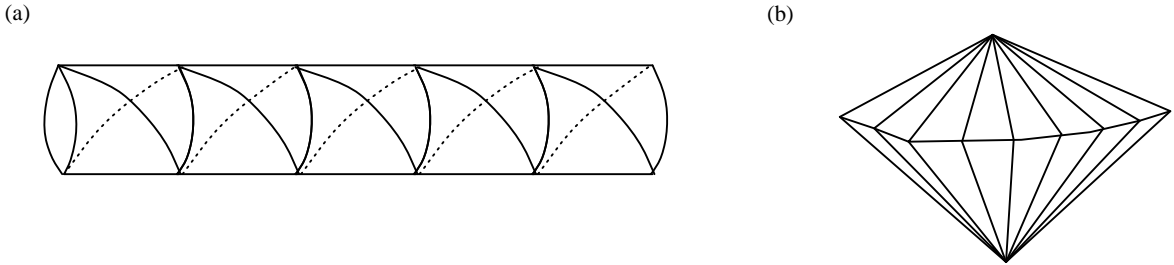}
\end{picture}
\end{figure}

For the $\phi^3$ graphs the high temperature ($\beta \to 0$) expansion
converges for $\tanh \beta < \half$, this implies that on the dual
triangulation the low temperature ($\beta \to \infty$) expansion
converges for $e^{-2\beta} < \half $ (ie for $\beta > \half \log 2
\approx 0.347$). Thus any critical value of $\beta$ must satisfy
$\beta_c \leq \half \log 2$, on the triangulated surface. In contrast
with $\phi^3$ graphs, there exist both graphs which never magnetize
and those which are permanently magnetized (see fig~\ref{fig:dual}).
 The argument
showing that there are no permanently magnetized $\phi^3$ graphs, can not
be used for the dual triangulation because the relation $\sum_i w_l(i,+)
< 2^l $ no longer holds. This is due to the fact that the vertices can
have coordination numbers greater than three (in fact the permanently
magnetized graph shown has points with coordination numbers of order 
$N$).

\subsection{Mechanisms of magnetization}
\label{sec:geom}
For this model the magnetization can be written as
$$M(p,\beta) = \frac{\sum\limits_G \frac{1}{\ssg} M_G (Z_G)^p}{\sum\limits_G
\frac{1}{\ssg} (Z_G)^p} , \eqq $$
where $M_G(\beta)$ is the magnetization for a given $N$-vertex graph
$G$, and we are taking the limit $N \to \infty$.

\noindent
In the limit $p \to 0$ we effectively have a quenched magnetization,
$$M_0(\beta) \equiv M(0,\beta)= \frac{1}{\gany} \sum_G \frac{1}{\sg}
M_G(\beta) , \eqq$$
where $\gany$ is the number of graphs with $N$ vertices in whichever
model is being looking at. This case has recently been investigated
numerically~\cite{BHJ}. The critical value for the magnetization
of this model will be denoted by $\beta_c$ and the critical coupling
constant for a given graph $G$ by $\beta_G^*$. One can show that the
behaviour of $M_0(\beta)$ at $\beta_c$ only depends on those graphs
which magnetize near $\beta_c$. We will illustrate this by
proving the result for the case in which graphs $G$ undergo a first
order phase transition; in this case only those graphs for which
$\beta_c - \epsilon < \beta_G^* \le \beta_c$, (where $\epsilon$
is an arbitrary small positive number) contribute to $M_0(\beta_c)$.
In actual fact the phase transitions of
individual graphs are second order, however the extension of the
result to this case is relatively straightforward and is omitted.
Now,
\eqtriv=\mycount
$$ M_0(\beta) = \frac{1}{\gany} \left[ \sum_{G:\beta_G^* \le \beta -
\epsilon} \! \! \!  M_G(\beta) \ + \sum_{G:\beta_G^* > \beta - \epsilon}
\! \! \! M_G(\beta) \ \right], \eqq $$
where the notation $G:\beta_G^* \le \beta - \epsilon$ means that we
are summing over graphs $G$ for which the inequality is satisfied and
we have absorbed the symmetry factors into the summations. Since the
system magnetizes at $\beta_c$,
\eqzero=\mycount
\def\inql{G:\beta_G^* \le \beta_c - \epsilon}
$$M_0(\beta_c -\epsilon) = \frac{1}{\gany} \sum_{\inql} \! \! \! M_G(\beta_c -
\epsilon) \ \ \mathop\to\limits_{\scriptscriptstyle N \to \infty} \  0 .\eqq $$
For a first order transition in which each graph jumps to $M_G^0$ at
its critical point,
$$\frac{1}{\gany} \sum_{\inql} \! \! \! M_G(\beta_c -\epsilon) \ \ge \
\frac{1}{\gany} \sum_{\inql} \! \! \! M_G^0 \ \ge 
\min(M_G^0) \
\frac{1}{\gany} \sum_{\inql} \! \! \! \! \! 1 \ \ . \eqq $$
While,
$$ \frac{1}{\gany} \sum_{\inql} \! \! \! M_G(\beta_c) \ \le
 \frac{1}{ \gany} \sum_{\inql} \! \! \! \! \! 1 \ \
\le \frac{1}{
\min(M_G^0)}
\frac{1}{\gany} \sum_{\inql} \! \! \! M_G(\beta_c - \epsilon) \ , \eqq $$
which tends to zero by (\number\eqzero).
Hence (\number\eqtriv) gives
$$M_0(\beta_c)= \frac{1}{\gany} \sum_{G:\beta_G^* > \beta_c -
\epsilon} \! \! \!  M_G(\beta_c) \  = \frac{1}{\gany} \sum_{G: \beta_c -
\epsilon < \beta_G^* \le \beta_c} \! \! \!  \! \! M_G(\beta_c) \ , \eqq $$
so that only graphs which magnetize near $\beta_c$ contribute to
$M_0(\beta_c)$. As it stands the proof cannot be used for
second order transitions (where $M_G^0=0$), but careful consideration
shows that in this case only graphs for which $\beta_c - \epsilon <
\beta_G^* \le \beta_c + \delta$ (where $\epsilon$, $\delta$ are small
positive numbers) can contribute to $M_0(\beta_c+\delta)$.
Thus the critical
exponents can only depend on the behaviour of graphs which are
magnetizing near the critical point $\beta_c$ and the magnetization is
as a result of spin-ordering of these graphs.

For non-zero $p$, a different geometric type of transition is possible
and this seems to be what occurs for large $p$. In this case the
magnetization is caused by the changing of the relative weights
between different graphs. This can best be understood by looking at a
simple model in which there are only two 
types of graph; suppose that there are
$n_1$ unmagnetized graphs ($n_1 \sim \exp f_1 N$) with partition
function $Z_1 \sim \exp \mu_1 N$ and $n_2$ magnetizable graphs ($n_2
\sim \exp f_2 N$) with magnetization $m(\beta)$ and partition
function $Z_2 \sim \exp \mu_2 N$. Then assuming that the magnetizable
graphs are the more numerous ($f_2>f_1$) and have smaller partition
functions ($\mu_2 < \mu_1$), which is certainly the case if we are
looking at transitions to tree graphs in model~\mone, the partition
function is
$$ Z_N(p) = n_1 {Z_1}^p + n_2 {Z_2}^p \similar e^{(f_1 + p \mu_1) N} +
e^{(f_2 + p \mu_2) N} \eqq $$
and the magnetization,
$$ M(p,\beta)
 \similar \frac{1}{Z_N(p)} \ m(\beta) \ e^{(f_2 + p \mu_2) N} .\eqq $$
In the thermodynamic limit the magnetization $M(p,\beta)$ is zero if $f_1+ p
\mu_1 > f_2 + p \mu_2$, that is, if $p (\mu_1 -\mu_2) >(f_2 - f_1)$.
As $\beta$ is varied the difference $\mu_1 - \mu_2$
changes and at the point for which the inequality no longer holds ($\beta_c$)
the magnetization jumps from zero to $m(\beta_c)$. Thus there is a
first order transition.

In the large $p$ limit for which
$\beta_c \to \infty$ it is possible in model~\mone~to have
magnetized graphs with $r \equiv n_1/N$ 
arbitrarily close to $\half$,
so that even if the
transition is generically first order the discontinuity,
$\Delta r$, could be zero in this
limit (ie perhaps $\Delta r \to 0$ as $p \to \infty$).
For $p \to \infty$ there is such a
transition, for model~\mone, between unmagnetized tree graphs and
magnetized non-tree-like graphs, with no discontinuity~\cite{Wex1,Wex2}.

\section{Conclusion}
\label{sec:conc}
\begin{figure}[htbp]
\caption{Possible phase diagrams}
\label{fig:phase}
\begin{picture}(100,140)(15,0)
\epsfbox{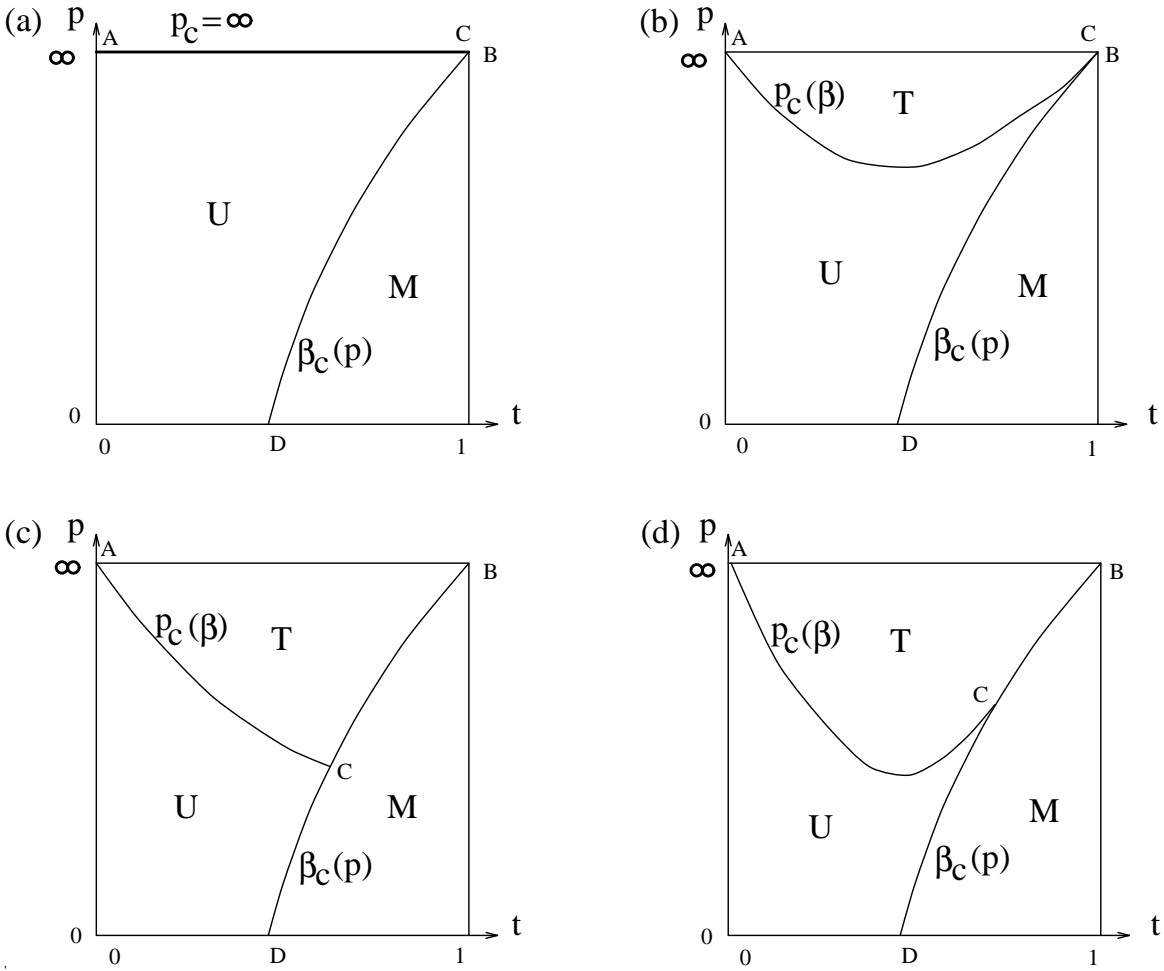}
\end{picture}
\end{figure}

In figure~\ref{fig:phase} we have drawn possible forms of the phase
diagram for model~\mone. The magnetized region is labelled M, the
tree-like region T and the remaining unmagnetized non-tree-like region
U. We would expect similar diagrams for models~\mtwo~and~\m3, but in
the latter case it is not clear what is happening in the T region,
as we have not identified the maximal graphs for intermediate values
of $\beta$.

For models~\mone~and~\mtwo, we know that the partition functions of
graphs increase as we replace closed loops with disconnected
structures. Since the graphs are becoming
less well-connected one might expect
that they tend to magnetize at higher values of $\beta$ and have
lower magnetizations at a given value of $\beta$. As $p$ is increased
in this model, the graphs with the larger partition functions make a
relatively larger contribution to $Z_N(p)$ and so we
expect that $\beta_c(p)$ will increase. There is some evidence for this
from numerical simulations~\cite{BaiJoh2,ADJT,BFHM,KowKrz}. We might
also expect the magnetization, for a fixed $\beta$, to decrease as $p$
is increased and again there is some evidence for this from
Monte Carlo simulations~\cite{ADJT}. It should of course be noted that
these simulations were for model~\m3, for which we do not know the
maximal graphs. However, since all the maximal graphs that we {\it have}
identified are not
well-connected and do not magnetize, it is quite likely that this also
applies for model~\m3 at intermediate values of $\beta$; it also
follows that in none of the models do we expect $\beta_c(p \! \to \!
\infty)$ to be finite.

In the papers of Wexler~\cite{Wex1,Wex2} the $p \to \infty$ limit of
model~\mone~is studied in the low temperature (large $\beta$)
expansion. In this limit there is a critical coupling $e^{-2\beta_c}
\sim 1/p$ below which the dominant graphs are tree-like being made up
of large numbers of almost disconnected baby universes within which
the spins are all aligned. This phase has $\gst=\half$ and our results
indicate that this behaviour must persist beyond the large $\beta$
expansion, in fact for all $\beta >0$. Above $\beta_c$ the dominant
graphs contain few baby universes each of whose volume (number of
vertices) is very large and $\gst=-\half$; 
thus it seems that the properties of the M region do
not depend strongly on p. At the critical point, where both the number
of baby universes and their volume diverge, it is found that
$\gst=\third$. This result is also found in the paper of Ambj{\o}rn et
al~\cite{ADJ} who considered a restricted model in which only phase
interfaces of minimum length are allowed; they obtain the result that
the exponent at the critical point $\gamma^{*}$ is related to that in
the large $\beta$ phase, $\og$, by $\gamma^{*} = \og / (\og -1)$ so
long as $\og<0$. This phase transition at $p \to \infty$ is
geometrical, as discussed in section~\ref{sec:geom}, rather than one
of spin alignment; it occurs because the partition functions of
magnetized graphs are catching up with the others. It seems highly
likely that this phenomenon will persist to finite $p$ although a
calculation of the $1/p$ corrections (or equivalently the contribution
of domain boundaries of length greater than two in the model
of~\cite{ADJ}) is needed to rule out the phase diagram of~\ref{fig:phase}b
analytically. 

The extent of the tree-region T is not yet fully understood. We know
that at $\beta=0$, where the Ising partition function is one, the
model is independent of $p$ and has $\gst=-\half$; in addition, the
absence of extra transitions in models with $c<1$ suggests that the line
AC separating the U and T regions does not go below $p=2$. We cannot
rule out the possibility that $p_c(\beta)=\infty$ as shown in
figure~\ref{fig:phase}a. However, the truncated models do display a
region of relatively rapid changeover to the T region and it seems
likely that the full models have a phase transition (with $p_c t
\approx 6$, $p_c t^3 \approx 8$ for models~\mone~and~\m3~respectively
at small enough $t$).

For model~\mone,
$x \equiv \half - r$ (where $r=n_1/N$ is the ratio of loops
of length one to vertices) serves as an order parameter. In the region T,
$x=0$ and approaching the line AC from below at fixed
$\beta$, we would expect $x$ to decrease continuously to zero, that is
a second (or higher) order phase transition.
This is because, just below the critical line,
the largest contributions to $Z_N(p)$ come from diagrams which are
tree-like except for a number of closed loops. As $p$ is increased the
number of closed loops (excluding loops of length one)
decreases to zero in a continuous fashion (see for
example fig~{\graphI}b, which shows how $\rbra$ approaches
 a half as $p$ is
varied, for the truncated model).

The transition to T from M across CB is rather different.
 Then we know that just below the
critical line the partition function is dominated by magnetized graphs
(which cannot be tree-like or almost tree-like, as such graphs do not
magnetize for $t \ne 1$) and such graphs have $r \ne \half$. Thus we
might expect there to be a first order transition with $r$ jumping
discontinuously. It is tempting to hypothesise that
for the line DC between the M and U regions there is a continuous
spin-ordering transition whose characteristics are determined 
by the graphs which are
actually magnetizing at the critical point
$\beta_c(p)$, as happens for the quenched case ($p \to 0$).
On the other hand the line CB between the T and M regions is
a geometrical phase transition in which the transition is caused
by the changes in geometry due to the changing relative weights
(ie ${Z_G}^p$) of the magnetized and tree-like graphs.

For model~\m3, there is direct
numerical evidence from computer simulations. In this case it is not
as clear what we should use as an order parameter for the U T
transition, but $n_5/N$ (the
ratio of closed loops of length five to vertices) seems a
plausible candidate. For both the fractal and ladder graphs this
quantity is zero and if, as seems possible, the intermediate maximal
graphs consist of mixtures of fractals arranged in a ladder-like
fashion, then it would be zero in the whole T region. Unfortunately,
none of the numerical studies looked at this quantity in detail.
Reference~\cite{ADJT} measure the quantity for $p=16$ (in fig 12 of
that paper), but only at the critical point $\beta_c$, where it is
very close to its pure gravity value.
However, references~\cite{BaiJoh1,BaiJoh2,ADJT} do give graphs of $n_3/N$ for
various values of $\beta$. This quantity peaks at some value of
$\beta$ below $\beta_c$ and drops to the pure gravity value as $\beta
\to 0$ or $\beta \to \beta_c$. The height of the peak grows with $p$
in an apparently linear fashion,
but we know that it is bounded by the relation $n_3/N \le \third$. It would be
interesting to see how this quantity saturates as $p$ is increased,
for $\beta$ near $\beta_c$, and to compare it with the results we have
from the truncated model in the limit of small $\beta$.

It is interesting to examine the average maximal radius, $r_{max}$,
of the ensemble. For a given graph
$r_{max}$ is defined in the following fashion. Take an arbitrary
point on the graph (called the ``centre''), mark all of its neighbours
as being at a distance one from the centre, mark all the unmarked
neighbours of these points as being at distance two
and so on. The maximal extension is then the distance of the furthest
point from the centre and averaging over different centres gives
$r_{max}$ on the graph. This procedure can be carried out on either
the \ph3 graph or the dual triangulation, giving two different
definitions, which we will denote $r_{max}^{G}$ and $r_{max}^{T}$
respectively. In figure~7 of~\cite{ADJT} $r_{max}^{T}$ is plotted for various
different models; the curve has a minimum for values of $\beta$ just
below $\beta_c$ and its depth increases with $p$ up to the maximum
value considered, $p=16$.
Both fractals and ladders have very small values of
$r_{max}^T$ and so as the T phase is
approached one would expect $r_{max}^T$ to fall.
Let $p^*$ be the value of $p$ at the point C in fig~\ref{fig:phase}c
and now suppose that $p$ is held fixed at $p<p^*$ and we approach the
magnetization transition from the U region; then as $\beta$ increases
we also get closer to the T region and thus might expect to see a
decrease in $r_{max}^T$, which disappears when the system magnetizes. As
$p \uparrow p^*$ this effect would become more pronounced until at $p$
just above $p^*$ we expect to see a sharp transition into a branched
polymer-like phase which immediately disappears as $\beta$ increases
and the system magnetizes; $r_{max}^T$ would decrease sharply as the
critical region around C is approached and then rebound when the
system magnetizes. This behaviour seems very like that observed
in~\cite{ADJT} which suggests to us that even the largest value of $p$
used in the simulation is probably no greater than $p^*$. If $p$ is
significantly greater than $p^*$ we would expect the minimum value of
$r_{max}^T$, when the ensemble is most branched polymer-like, to occur
somewhere in the middle of the T region rather than at the
magnetization transition; so as $p$ increases beyond $p^*$ the minimum
of $r_{max}^T$ will move away from the magnetization transition. It is
difficult to be entirely certain in interpreting the behaviour of
$r_{max}^T$ without knowing how the minimum value scales with $N$; it
would be useful to have simulation data on this and also at much
larger $p$ to be certain of our interpretation.

Although we have worked entirely with Ising models many of our
considerations probably extend to other spin systems which have second
order phase transitions on regular lattices; in particular the maximal
graphs and the general behaviour of the truncated models are likely to
be the same.

\bigskip

We would like to thank Th\'ordur J\'onsson for giving us advanced details
of~\cite{ADJ} and to acknowledge the support of the SERC under grant GR/J21354
and through the research studentship GR/H01243.

\appendix
\section{Proof that ring graphs are maximal} 
\label{app:ring}

In order to complete the proof that rings are maximal for model~\mtwo,
we need to show that each of the subgraphs listed in
section~\ref{sect:ring}~can be
eliminated in a way that increases the partition function.

Consider the case for the 2-loop first. A propagator dressed with $n$
2-loops and joining spins $S_a$ to $S_b$ (fig~\ref{fig:dress}c), 
has a contribution of
$$ G^{(n)}(S_a,S_b)= 2^{2n} C^{3n+1} \left( (1+t^2)^n + S_a S_b t^{2n+1}2^n
\right) = C' (1+ t' S_a S_b), \eqq $$
where $C=\cosh \beta$, $t=\tanh \beta$ and
$$C'=2^{2n}C^{3n+1}(1+t^2)^n \eqq$$
$$ t'= \left( \frac{2 t^2}{1+t^2} \right)^n t \equiv Y^n t . \eqq $$
Thus dressing a propagator effectively renormalizes the coupling
constant for that link. The partition function for the whole of the
original graph is given by
$$ \sum_{S_a S_b} Z(S_a,S_b) \ Z_1(S_a,S_b), \eqq $$
where $Z(S_a,S_b)$ is the partition function of the remainder of the graph
($Z(S_a,S_b) \ge 0$) and $Z_1$ is that for the subgraph being
replaced (fig~\ref{fig:loop}a).
The partition function for the graph after the
replacement has $Z_1$ replaced by $Z_2$
(fig~\ref{fig:loop}c). The powers of two and the factors of $C$ and
$C'$ are the same for both graphs and cancel with terms in
$Z_0^{-1}$, so that we can drop these factors
which gives,

\setcounter{equation}{\mycount}
\addtocounter{equation}{-1}
\global\advance\mycount by 2
\vbox{
\begin{eqnarray}
 Z_1 & = & 1 + Y^{n+m} t^2 + S_a S_b t^3 \left(Y^n + Y^m\right), \\
 Z_2 & = & 1 + t^2 + S_a S_b \left(2 t^3 Y^{n+m}\right) .
\end{eqnarray}}

\noindent Hence,

\setcounter{equation}{\mycount}
\addtocounter{equation}{-1}
\global\advance\mycount by 2
\vbox{
\begin{eqnarray}
 \Delta Z &=& t^2 \left( 1 - Y^{n+m} + S_a S_b t \left( 2 Y^{n+m} - Y^{n}
-Y^{m} \right) \right) \\
          &=& t^2 \left( \left(1-Y^n \right) \left( 1 - S_a S_b tY^m \right)
+ Y^n \left(1- Y^m \right) \left( 1 - S_a S_b t\right) \right) \ge 0 .
\end{eqnarray}}

\noindent Noting that $0\le Y,t \le 1$ and $n,m \ge 0$, we can see
that $\Delta Z$ is not negative and that $\Delta Z=0$ occurs for $t=0$
and $t=1$, as we would expect. The partition function of the graph is
increased by the replacement in fig~\ref{fig:loop}.

The triangular case (fig~\ref{fig:tri}) can be proven in the same way
as the 2-loop case. However, for the square and pentagon, the
expression for $\Delta Z$ is so complicated that it is extremely
difficult to regroup the terms and write $\Delta Z$ in a form which is
manifestly non-negative. An inductive method can be used to prove each
of the cases and we will indicate below how one would use it to prove
the case of the square (fig~\ref{fig:squ}). For this case we need to
prove that $\Delta Z_{klmn}(S_a,S_b,S_c,S_d)$ where

\vbox{
$$ \Delta Z_{klmn} t^{-2} \qquad = \qquad 2 + 
t \left( S_aS_b Y^k\left(2 Y^n-1\right) +
S_cS_d Y^m\left(2Y^l-1\right) -S_bS_cY^l -S_dS_aY^n \right) + \quad $$
$$ t^2 \left( 1- Y^{k+l+m+n} -
S_aS_c\left(Y^{k+l}+Y^{n+m}\right) -
S_bS_d\left(Y^{m+l}+Y^{n+k}\right) \right) +$$
$$ t^3 \left(S_aS_b Y^n \left(2 Y^k - Y^{l+m} \right)
 + S_c S_d Y^l \left( 2 Y^m
- Y^{n+k} \right) - S_bS_c Y^{k+n+m} - S_dS_a Y^{m+l+k} \right) + $$
$$ t^4 \ S_aS_bS_cS_d \left( 4 Y^{k+l+m+n} - Y^{k+m} -Y^{n+l} \right)
, \eqq $$}

\noindent is positive for all $0<t<1$,
and all values of the spins, $S_a,S_b,S_c,S_d$.
It is easy to prove that $\Delta Z_{0 0 0 0} > 0$ for $0<t<1$ and any
values of the spins $S_a,S_b,S_c,S_d$. We define a difference operator $D_k$
by
$$D_k \Delta Z_{klmn} = \Delta Z_{ (k+1) l m n} - Y \Delta Z_{klmn}
\eqq $$
and similarly for $D_l$, $D_m$, $D_n$. Now, $\Delta Z_{klmn}$ has the
form,
$$ \Delta Z_{klmn}= A+ Y^k B , \eqq $$
where $A$, $B$ are independent of $k$, but in general depend on all
the other variables. Hence,
$$ D_k \Delta Z_{klmn} = (1-Y) A, \eqq $$
which is independent of $k$; note that $(1-Y)$ is a positive factor.
Now,
$$ \Delta Z_{(k+1) lmn} = Y \Delta Z_{klmn} + D_k \Delta Z_{klmn}.
\eqq $$
If we can show that $D_k \Delta Z_{klmn} > 0$ for any $l$, $m$, $n$
and similarly for $D_l$, $D_m$ and $D_n$ acting on $\Delta Z$, then,
since $\Delta Z_{0000} > 0$, the result
$\Delta Z_{klmn} > 0$ for all $k$, $l$, $m$, $n \ge 0$ follows by induction.

In order to prove that $D_k \Delta Z_{klmn} > 0$ we use the fact that
the difference operators commute so that,
$$ D_k \Delta Z_{k l (m+1) n} = Y (D_k \Delta Z_{klmn}) + D_m ( D_k
\Delta Z_{klmn}) . \eqq $$
Now we need to prove that $D_k \Delta Z_{k 0 0 0} > 0$ and that $D_m
D_k \Delta Z_{klmn} > 0$ (and similarly for all pairs of difference
operators). Proceeding in this way the problem is reduced to showing that
 the following
quantities are positive: $\Delta Z_{0000}$, $D_k \Delta Z_{k000}$
(similarly for $D_l$ etc), $D_k D_l \Delta Z_{k l 0 0}$ (similarly for
all possible pairs), $D_k D_l D_m \Delta Z_{k l m 0}$ (all possible triples) 
and $D_k D_l D_m D_n \Delta Z_{klmn}$. Each of these expressions
depends only on $t$ and the spins, and is thus relatively easy to
prove (especially if a computer is used to check each case). Thus
finally we have $\Delta Z_{klmn} > 0$ for $0<t<1$. The pentagonal case
can be proved in the same fashion.

\section{Derivation of recurrence relations} 
\subsection{Model~\m3}
\label{app:tri}
\begin{figure}[htb]
\caption{Replacement does not change $n_3$}
\label{fig:1}
\begin{picture}(70,35)(-20,0)
\epsfbox{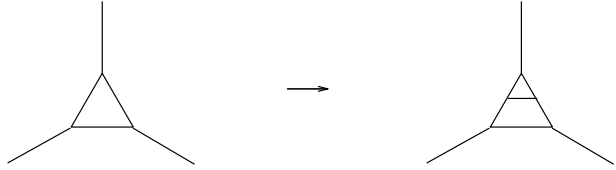}
\end{picture}
\end{figure}

The recurrence relation (\number\rra) for two-particle
irreducible planar graphs is derived by considering
the effect of replacing a point P, on a graph $G$, by a triangle. Such a
replacement increases the number of vertices in the graph by
two and increases the number of triangular loops ($n_3$) by either one
(fig~\ref{fig:fractet}a) or zero (fig~\ref{fig:1})
depending on whether or not the point P of graph $G$
was lying on a triangle. Note that since self-energy terms have
been eliminated the only possible diagram for which P could lie on two
triangles is given by fig~\ref{fig:fractet}b,
which has $N=4$ vertices; we shall ignore
this case and derive a recurrence relation valid for $ N \geq 6$.

Graphs with $N+2$ vertices and $n_3+1$ triangles can be made from
those with $N$ vertices and $n_3$ triangles by replacing one of $N-3n_3$
points (fig~\ref{fig:fractet}a), or from graphs with $N$ vertices and $n_3+1$
triangles by replacing one of $3(n_3+1)$ points (fig~\ref{fig:1}). We will
refer to the points which yield the correct number of triangles when
replaced (ie $n_3+1$) as being `allowed' points. Clearly all such
graphs can be made by this method.

We need to determine the effect of such a
replacement on the symmetry factor of a graph. Suppose that
graph $G$ has symmetry group $\cal G$ of order $s$. Since $G$ is planar
we can think of it as being a polyhedron and the members of $\cal G$
as being rotations. By making rotations we can interchange a point
P with any of a set R (of size $f_1$) of equivalent points.
Now, rotations that leave P in the same position give a subgroup $\cal
H$ of $\cal G$ which has order ${s}/{f_1}$ (which will equal
either one or three, since there are three lines emerging from each
vertex). Replacing P by a triangle T, we group the triangles of
the new graph $G'$ into equivalence classes. If T is a member of a class
of size $f_2$ then the order of the symmetry group of $G'$ is $s' =
 s {f_2}/{f_1}$.

If we consider the equivalence classes of all the allowed points in
all graphs with $N$ vertices, then there is a one-to-one correspondence
between these classes and the classes of triangles in all the graphs
with N+2 vertices. A replacement of a point P by a triangle T defines
a mapping between classes of points and of triangles. Let us associate
a weight ${f_1}/{s}= {f_2}/{s'}$ for this mapping between
classes. That is, we are associating a weight of ${1}/{s}$ for each
allowed point in a graph that is being mapped from, or equivalently a weight
of ${1}/{s'}$ for each triangle in a graph that is being mapped to.
The total weight for allowed points is given by,
$$ \sum_{graph \ i} \sum_{allowed \atop points} \frac{1}{s_i}
= \sum_{graph \ i \atop (N,n_3)} (N-3 n_3) \frac{1}{s_i} + 
\sum_{graph \ j \atop (N,n_3+1)} 3(n_3+1) \frac{1}{s_j} \eqq $$
$$ = (N-3n_3) \, \g3 (N,n_3) + 3(n_3+1) \, \g3 (N,n_3+1) .$$
However, this is equal to the weight for the graphs being mapped to,
due to the one-to-one correspondence, which is
$$ \sum_{graph \ i \atop (N+2,n_3+1)} \sum_{triangles} \frac{1}{s_i'} =
\sum_{graph \ i} (n_3+1) \frac{1}{s_i'} = (n_3+1) \, \g3 (N+2,n_3+1) . \eqq
$$
Equating these gives the claimed result.

\subsection{Model~\mone} 
\label{app:tad}
\begin{figure}[htb]
\caption{(a) Replacement (b) Replacement on tadpole (c)
Orientated version}
\label{fig:2}
\begin{picture}(100,40)(22,0)
\epsfbox{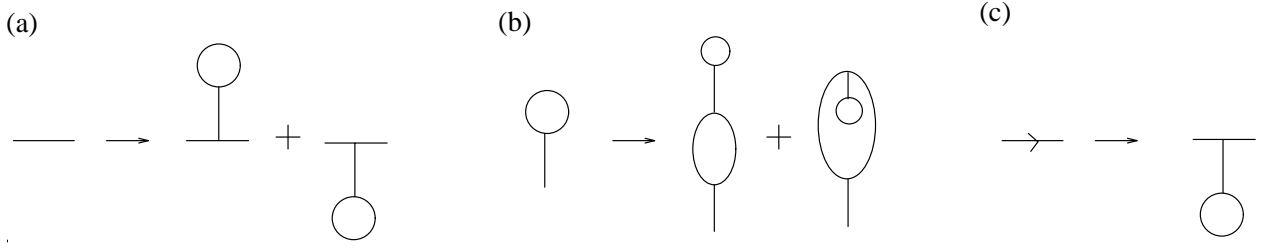}
\end{picture}
\end{figure}

The recurrence relation (\number\rrb) for
model~\mone~can be derived in a similar fashion, by considering the addition
of a tadpole to a link (see fig~\ref{fig:2}a). There is the added complication
that the tadpole can be orientated in one of two different directions,
(we distinguish between these configurations because they are
counted separately in the matrix model). We will consider the
creation of graphs with $N+2$ vertices and $n_1+1$ tadpoles, from $N$ point
graphs with either $n_1$ tadpoles (see fig~\ref{fig:2}a) or $n_1+1$ tadpoles
(fig~\ref{fig:2}b). For the $N+2$ vertex graphs we will group the tadpoles into
equivalence classes. However, there are no rotations that leave a
given tadpole fixed (since this would correspond to rotating a
polyhedron, whilst keeping an edge and an adjacent face fixed, which
is impossible). Thus, rotations just interchange the tadpoles within
each equivalence class in a cyclic fashion and the order of the
symmetry group, $s'$, is just equal to the size of the equivalence
classes (all of them being the same size).
(Incidentally, this means that the order of the symmetry group of a
graph divides the number of tadpoles. Hence,
if the number of tadpoles is prime, $p$, then the graph has a
symmetry factor of either 1 or ${1}/{p}$.)

For the $N$ point  graphs, we decompose the graphs into sets of
allowed links (that is, links which when we add a tadpole to them
give us a total of $n_1+1$ tadpoles).
Each $N$ point graph has $\frac{3N}{2}$ links and $3N$
orientated links (ie we count each link twice; once with an arrow
going in one direction and once with the arrow reversed).
The orientated links are collected into classes which contain links
that can be rotated into each other. A link may or may not be in the
same class as its reversed version. It is impossible to make a
rotation that leaves a link in the same position and with the same
orientation. As a result, we again have that the equivalence
classes are all of the same size, which is equal to the order of the
symmetry group, $s$. Now, the addition to an orientated link (as in
fig~\ref{fig:2}c) of a tadpole on the right-hand side, defines a
mapping between
equivalence classes of allowed links in $N$ point graphs and of
tadpoles in $N+2$ vertex graphs. This mapping is again one-to-one.
The sizes
of the equivalence classes equal the orders of the symmetry groups, so
that the number of classes for a graph equals the number of allowed
links divided by the order of the group. Equating the numbers of
equivalence classes, we get,
$$\sum_{graph \ i \atop (N,n_1)} 2 \left(\frac{3N}{2} - n_1 \right)
\frac{1}{s_i}  + \sum_{graph \ j \atop (N,n_1+1)} 2 (n_1 + 1)
\frac{1}{s_j} = \sum_{graph \ k \atop (N+2,n_1+1)} (n_1+1)
\frac{1}{s_k} \eqq $$
$$ 2 \left( \frac{3N}{2} - n_1 \right) \ggone(N,n_1) + 2 (n_1+1)
 \, \ggone(N,n_1+1)
= (n_1+1) \, \ggone(N+2,n_1+1) , \eqq $$
which gives the claimed result.


\end{document}